\documentclass[conference, 10pt]{IEEEtran}
\IEEEoverridecommandlockouts
\usepackage[compress]{cite}

\usepackage{subfigure}
\usepackage{float}

\usepackage[bookmarks=false, hidelinks]{hyperref}

\usepackage{amsmath, amssymb, amsfonts}
\allowdisplaybreaks
\usepackage{bm}                         % Bold calligraphic letters
\usepackage{multirow}
\usepackage{cases}                       % for numcases environment

\usepackage[normalem]{ulem}             % strike text

\usepackage{algorithmic}
\usepackage{graphicx}
\usepackage{textcomp}
\usepackage{xcolor}
\def\BibTeX{{\rm B\kern-.05em{\sc i\kern-.025em b}\kern-.08em
    T\kern-.1667em\lower.7ex\hbox{E}\kern-.125emX}}

\usepackage{balance}

\begin{document}

\title{Modelling multi-cell edge video analytics
\thanks{This work has been submitted to IEEE for publication. Copyright may be transferred without notice, after which this version may no longer be accessible.}
}
%The authors are with the School of Electrical Engineering and Computer Science, KTH Royal Institute of Technology, Stockholm, Sweden.\\(e-mail: jaumeap@kth.se; vfodor@kth.se)
%Details\\
%Location\\
%email@institution.com

%\author{Jaume Anguera Peris \orcidicon{0000-0003-2817-7257}, \IEEEmembership{Student Member, IEEE}, Viktoria Fodor \orcidicon{0000-0002-2764-8099}, \IEEEmembership{Member, IEEE}
%Authors
%}

\author{\IEEEauthorblockN{Jaume Anguera Peris and Viktoria Fodor} %{(\color{blue}Last edited: \today})}
\IEEEauthorblockA{School of Electrical Engineering and Computer Science\\
KTH Royal Institute of Technology, Stockholm, Sweden\\
Email: \{jaumeap,vfodor\}@kth.se}
%\and
%\IEEEauthorblockN{Viktoria Fodor}
%\IEEEauthorblockA{School of Electrical Engineering and Computer Science\\
%KTH Royal Institute of Technology\\
%Stockholm, Sweden\\
%Email: vfodor@kth.se}
}

\maketitle

\begin{abstract}
Edge intelligence is a scalable solution for analyzing distributed data, but it cannot provide reliable services in large-scale cellular networks unless the inherent aspects of fading and interference are also taken into consideration. In this paper, we present the first mathematical framework for modelling edge video analytics in multi-cell cellular systems. We derive the expressions for the coverage probability, the ergodic capacity, the probability of successfully completing the video analytics within a target delay requirement, and the effective frame rate. We also analyze the effect of the system parameters on the accuracy of the detection algorithm, the supported frame rate at the edge server, and the system fairness.
\end{abstract}

\begin{IEEEkeywords}
Edge intelligence, stochastic geometry, queuing theory, edge computing
\end{IEEEkeywords}

% Possible titles - please add more
% A tractable approach for video edge analytics in multi-cell cellular networks
% Tractable modelling of video edge analytics in multi-cell wireless networks
% Performance and fairness of video edge analytics in multi-cell wireless networks
% Tractable modelling of multi-cell edge-intelligent systems for video analytics

%\VF{Comment: we have a bit of mess of video analytics, image processing, object detection .... We need to harmonize it throughout the paper. Whatever expression we use in the title, it needs to show up in the abstract and in the introduction.}

\section{Introduction}
\label{sec:introduction}
The evolution of mobile devices has introduced tremendous innovation opportunities for developing new services beyond personal communications. These advances have brought new mobile applications, and with them, the proliferation of large amounts of app-level data. At the same time, we have experienced major breakthroughs in deep learning, which have made it possible to manage and leverage the rise in data volumes for solving complex problems, such as image classification, language processing, or intrusion detection \cite{chen2019deep}.

% \JP{However,} as we strive for more accurate and reliable deep learning models, the computationally and energy-constrained mobile devices make real-time processing intractable. One prominent solution to extend the capabilities of mobile devices is to %offload computationally intensive tasks to a 
% use a centralized cloud infrastructure \cite{saeik2021task}, but this solution also introduces new challenges because it incurs large communicating delays and enormous network traffic, and raises concerns about scalability and privacy \cite{chen2019deep}. To address these issues, edge computing offers a distributed solution that brings the computational resources closer to the mobile devices, thus providing low latency, energy efficiency, privacy protection, and reduced bandwidth consumption \cite{zhou2019edge}. Edge computing also makes it easier to gather richer data and learn from the behaviour of nearby mobile devices, making it possible to explore deep learning models for interactive cloud gaming, cognitive assistance, and real-time video analytics \cite{zhou2019edge}.

However, as we strive for more accurate and reliable deep learning models, the computationally and energy-constrained mobile devices make real-time processing intractable. One prominent solution to extend the capabilities of mobile devices is to offload the computationally intensive tasks to an edge computing server. Edge computing offers a distributed solution that brings the computational resources closer to the mobile devices, and provides lower latency, more energy efficiency, better privacy protection, and lesser bandwidth consumption than its counterpart cloud-based infrastructure \cite{zhou2019edge}. Edge computing also makes it easier to gather richer data and learn from the behaviour of nearby mobile devices, making it possible to explore deep learning models for interactive cloud gaming, cognitive assistance, and real-time video analytics \cite{zhou2019edge}.

Indeed, the growing interest in both edge computing and artificial intelligence has coined the term \textit{edge intelligence}. This area of research is however in its early stages, and most of the proposed solutions are optimized to be deployed on local networks, such that it can be assumed that all users have similar network conditions and similar probabilities of successfully completing their task. In large-scale cellular networks, aspects of noise, fading, or interference can significantly impact the wireless communication \cite{saeik2021task}, so those assumptions no longer hold, and analyzing the communication between the mobile users and the edge server becomes as relevant as analyzing the implementation of the deep learning models. Hence, there needs to be a tractable approach to understand how the network, the computing resources, and the deep learning models affect altogether the task offloading problem.

% Recent works have shown that this is specially a problem for wireless communications because mobile users operate under varying network and traffic conditions, which can negatively effect the performance of deep learning algorithms for real-time applications \cite{ran2018deepdecision, galanopoulos2020measurement}. To solve this problem, stochastic geometry offers an analytical tool for modelling wireless networks for edge computing in latency-stringent applications \cite{ko2018wireless}, and queuing theory offers a tractable approach for modelling the traffic dynamics associated with data stream processing applications \cite{da2021latency}. 

In this work, we present a mathematical framework for analyzing edge-intelligent, multi-cell cellular systems supporting video analytics. We specifically leverage the tools of stochastic geometry, queuing theory, and deep learning for modelling the entire offloading process under different accuracy-latency constraints. We provide a thorough analysis of the relationship between the parameters of the video analytics, the uplink transmission, the edge server, and the deep learning algorithm, and evaluate their trade-offs via numerical results. Overall, our discussions and results provide the first steps towards designing efficient resource-allocation strategies and traffic-control protocols for edge-intelligent, multi-cell cellular systems.

The remainder of this paper is organized as follows. Section \ref{sec:relatedwork} reviews the related work. Section \ref{sec:system_model} describes the system model for the entire offloading process. In Section \ref{sec:system_analysis}, we derive the expressions for the performance metrics that evaluate the trade-offs between the different parameters in the system. Then, we provide the numerical results in Section \ref{sec:numerical_results}, followed by the concluding remarks in Section \ref{sec:discussion}.

\section{Related work}
\label{sec:relatedwork}
Edge intelligence is a promising solution to assist mobile devices with computationally-intensive and latency-stringent applications, and it is recently receiving a lot of attention for visual-aid services, e.g., augmented reality \cite{liu2018dare} and video edge analytics \cite{ran2018deepdecision}. As a response to these emerging interests, recent works have built proof-of-concept testbeds 
%for analyzing the compatibility of edge-intelligent wireless systems for supporting those services
\cite{liu2018dare, ran2018deepdecision, galanopoulos2020measurement}. DARE \cite{liu2018dare} focuses on the edge server and analyzes the effect of the frame size, the frame rate, and the deep learning model on the task offloading problem. DeepDecision \cite{ran2018deepdecision} considers the same parameters but instead focuses on the mobile users and the wireless network. The work in \cite{galanopoulos2020measurement} 
%extends the results from those previous works and 
uses its own measurements to design data-driven statistical models for real-time inference. 

The results from the above works highlight the suitability of edge intelligence for improving the accuracy and latency of visual-aid services, but their simplified models for the communication between the mobile users and the edge servers do not capture the inherent aspects of large-scale cellular networks. 
In this context, \cite{wang2020joint} models the performance of edge-intelligent systems with parallel processing for video analytics, and \cite{ko2018wireless} models the performance of edge computing in multi-cell networks, but does not address the performance of the application itself.

%\JP{The work in \cite{wang2020joint} analyzes the performance of video analytics in a single-cell edge server, and \cite{ko2018wireless} analyzes the performance of edge computing in multi-cell cellular networks, but does not address the performance of the application itself.}
% \VF{The performance of video analytics in single cell edge computing is modeled in \cite{wang2020joint}, while \cite{ko2018wireless} considers edge computing performance in multi-cell networks, however, without addressing the application performance itself.}

%In this regard, prior works have motivated the use of edge-intelligent cellular networks for visual-aid services \cite{liu2019virtualedge}, as well as the use of stochastic geometry for analyzing the uplink transmission in multi-cell cellular networks with edge computing \cite{ko2018wireless}.

Our work extends the above results and presents a mathematical framework for analyzing how the conditions of the network, the frame size, and the latency requirements affect the transmission rate, the achievable frame rate, the probability of successfully completing the video analytics, and the accuracy of the detection algorithm.

\section{System model}
\label{sec:system_model}
\subsection{Network}

%Consider a multi-cell wireless network where all the base stations employ frequency-division multiple access (FDMA) with reuse factor $\delta$, such that mobile users experience inter-cell interference but do not experience any intra-cell interference. \JP{Each mobile user connects to the base station that provides the maximum received power averaged over fading, and each base station has only one active mobile user scheduled on a given frequency band, which is arbitrarily chosen from all the mobile users located in its Voronoi cell.}
Consider a multi-cell wireless network where all the base stations employ frequency-division multiple access with reuse factor $\delta$, such that mobile users experience inter-cell interference but no intra-cell interference. Each mobile user connects to the base station that provides the maximum received power averaged over fading, and each base station has only one active user in a given frequency band, which is arbitrarily chosen from all the mobile users located in its Voronoi cell. Motivated by \cite{wang2014distance}, the locations of the base stations are modelled according to a two-dimensional homogeneous Poisson point process (PPP) with intensity $\lambda_b$. Similarly, the locations of the mobile users using the same frequency resources are considered to be independent and modelled according to some other stationary PPP with intensity $\lambda_u$.  Considering this, it can be shown that the distance between any user and its serving base station, denoted by $r$, is Rayleigh distributed. The proof follows from the null probability of a two-dimensional PPP \cite{chiu2013stochastic},
% in a circle of area $\pi r^2$,}
%Then, if we consider  %\sout{it is reasonable to assume that any user located at a distance $r$ from its serving base station has no other base station closer than $r$. Considering this,} the complementary cumulative density function of the distance \VF{between any user and its serving base station, noted by} $r$ can be derived from the null probability of a two-dimensional Poisson process in a circle of area $\pi r^2$,
\begin{equation*}
    \mathcal{P}[R>r] = \mathcal{P}\left[\text{no BS in a circle of area } \pi r^2\right] = e^{-\lambda_b \pi r^2},
\end{equation*}
from which the probability density function (PDF) is
\begin{equation}
    f_R(r) = 2 \lambda_b \pi r \, e^{-\lambda_b \pi r^2},\quad r\geq 0.
    \label{eq:pdf_rayleighDistribution}
\end{equation}
%Hence, the distance between any user and its serving base station follows a Rayleigh distribution, which is a well-known classical result \cite{wang2014distance}. %\sout{that describes the distribution of real cellular networks} 

%Since there is inter-cell interference, any base station receives the signal from its desired user as well as signals from other users using the same frequency resources.
For the rest of the paper, we consider a target user $k$ and its serving base station, and refer to the set of interfering mobile users as $\mathcal{Z}$. We denote the distances from the interfering users $z\in\mathcal{Z}$ to the base station of interest by $d_z$, and the distances from the interfering users to their serving base stations by $r_z$. Note that the distances $\{r_z\}_{z\in\mathcal{Z}}$ are identically distributed but not necessarily independent. The dependence comes from the structure of the Poisson-Voronoi tessellation and the restriction that there is only one active user in each frequency band. However, \cite{novlan2013analytical} shows that this dependence is weak and the distances $\{r_z\}_{z\in\mathcal{Z}}$ can be assumed to be independent and identically distributed (i.i.d.). Furthermore, the marginal distribution of $r_z$ can be assumed to be Rayleigh distributed as in \eqref{eq:pdf_rayleighDistribution}, which is motivated by the irregular deployment of base stations and the same null-probability argument as for $r$.

%\VF{With this assumption, the distances $r_z$ are independent and Rayleigh distributed with a probability density function $f_{R_z}(r_z)$ as in \eqref{eq:pdf_rayleighDistribution}.} 

%\sout{In particular, motivated by the study of an irregular deployment of base stations and the same reasoning as for deriving the distribution of $r$, we consider the distances $r_z$ to be independent and Rayleigh distributed and to have a probability density function $f_{R_z}(r_z)$ as in \eqref{eq:pdf_rayleighDistribution}.} \JP{(The result in \cite{novlan2013analytical} shows that the distances $r_z$ are independent of each other, but that does not tell us anything about the distribution of $r_z$, nor that it has to be the same as $f_R(r)$. We still need to motivate why are we choosing a Rayleigh distribution rather than any other distribution (e.g., uniform or log-normal)).}
% \begin{equation*}
%     f_{R_z}(r_z) = 2 \lambda_b \pi r_z \, e^{-\lambda_b \pi r_z^2},\quad r_z\geq 0,\, \forall z\in \mathcal{Z}.
% \end{equation*}

\subsection{Uplink transmission}
\label{sec:transmission}
%In wireless systems that support video edge analytics, 
In edge video analytics, the amount of data transmitted in the uplink, i.e., from the users to the base stations, is much larger than the amount of data transmitted in the downlink. Therefore, we focus on modelling the uplink transmission.
%\VF{In video edge analytics applications large amount of data is transmitted uplink, from the users to the base stations, and comparably little data downlink. Therefore, we concentrate on the modelling of the uplink transmissions.} 

We consider that %the signals transmitted from the mobile users to the base stations
the transmitted signals attenuate with path-loss exponent $\alpha > 2$ and experience i.i.d.\@ Rayleigh fading.
All the users utilize a maximum-ratio transmission precoding vector, such that the effect of the channel can be modelled 
%We model the effect of the channel
at the receiver by the multiplicative Exponential random variable $|g|^2\sim\exp(1)$. We define the largest possible time-frequency space within which the effect of the channel can be modelled as a scalar multiplication by $|g|^2$ as the coherence interval. We denote the coherence interval by $\tau_c = T_{coh}\, B_{coh}$, where $T_{coh}$ represents the coherence time, and $B_{coh}$ represents coherence bandwidth. Considering this, we define the number of coherence intervals that fit into a time-frequency block of $t$ seconds and $B$ Hertz as $N_{\textit{coh}}= t\, B/\tau_c$. %$N_{\textit{coh}}=\left\lceil t\, B/\tau_c \right\rceil$. %., where $\lceil\cdot\rceil$ is the ceil function. 
% \begin{equation*}
%     N_{\textit{coh}}=\frac{t\times B}{\tau_c} = \frac{t\times B}{T_{coh}\times B_{coh}}.
% \end{equation*}

Moreover, we consider that all mobile users utilize distance-proportional fractional power control of the form $r^{\alpha\epsilon}$, where $\epsilon\in[0,1]$ is the power control factor. We define a reference power at a distance of $1$ kilometer from the base station as $P$ and assume it to be the same for all the users. To model the maximum transmission power of the mobile users, we consider the transmit powers $Pr^{\alpha\epsilon}$ to be limited by the peak power $\Bar{P}$. Finally, we consider the noise power at the receiver to be $\sigma^2$. Under this transmission model, the signal-to-interference-noise ratio (SINR) for any randomly selected user $k$ at its serving base station is
\begin{equation}
    \text{SINR}_k = \frac{|g_k|^2 \,\ell(r,\alpha,\epsilon)\,r^{-\alpha}}{\sigma^2_k +\sum_{z\in\mathcal{Z}} |g_z|^2\, \ell(r_z,\alpha,\epsilon)\, d_z^{-\alpha}},
    \label{eq:SINR}
\end{equation}
where
\begin{equation}
    \ell(r,\alpha,\epsilon) = \min\left(Pr^{\alpha\epsilon},\Bar{P}\right), \label{eq:power_constrain}
\end{equation}
and the sub-indices $k$ and $z$ indicate if a variable depends on the location and the resources assigned to the target user $k$ or to the interfering users $z\in\mathcal{Z}$, respectively. 

For the uplink transmission, we consider that each base station assigns a fixed bandwidth $B_k$ to every active user within its Voronoi cell. Once a user $k$ has been assigned a frequency band, it estimates the state of the channel in each coherence interval from pilot signals and adjusts its coding and modulation according to its instantaneous SINR$_k$. As a result, the transmission rate $R_k$ is a random variable that depends on the user's location, the conditions of the channel, and the inter-cell interference.

\subsection{Mobile edge computing server}
%All the base stations in the network are equipped with an independent edge computing server that serves all the \VF{video analytics users within the Voronoi cell of the base station. We consider that there is a set of $K$ active video analytic users in the cell, denoted by $k$, for $k=1,\dots,K$, that are capable of continuously recording images and sending them over the wireless channel.}
%Within the Voronoi cell of each base station, we assume there is a set of $K$ active users, denoted by $k$, for $k=1,\dots,K$, that are capable of continuously recording images and sending them over the wireless channel.
All the base stations in the network are equipped with an independent mobile edge computing server that is responsible for serving all the video analytic users within the Voronoi cell of its corresponding base station. For our base station of interest, we consider that there is a set of $K$ active video analytic users, denoted by $k$, for $k=1,\dots,K$, that are capable of continuously recording images and sending them over the wireless channel. Once the images of these users are completely received at the base station, they are delivered to the server without delay and queued at an infinite buffer for being processed on a first-come-first-served basis. 

From the point of view of the server, each user $k$ is seen as an independent entity that offloads images to the server at an arrival rate of $\lambda_k$ frames per second. Altogether, since the communication system employs frequency-division multiple access, 
%all the users within the same cell are seen as $K$ parallel streams of asynchronous arrivals, 
the queue at the server observes the superposition of $K$ independent arrival processes with an aggregate arrival rate of $\lambda =\sum_k \lambda_k$ frames per second. Under these assumptions, the arrival process behaves as a Poisson process \cite{sriram1986characterizing}. Once in queue, the images are processed at the server in a deterministic amount of time $T_s$, which is determined by the resolution of the transmitted images and the neural network considered for the video analytics.

%When the images of all those users are queued at the server, 
%the queue observes the superposition of independent arrival processes with an aggregate arrival rate of $\lambda =\sum_k \lambda_k$ frames per second. Under these assumptions, the arrival process can be modelled as a Poisson process \cite{sriram1986characterizing}. 

%\VF{The images are processed in a} deterministic amount of time $T_{s}$, which is determined by the resolution of the transmitted images and the neural network considered for the video analytics.
% \VF{and the applied image processing algorithm}. \sout{Since this is the first approach to our proposed model,}

In this paper, we consider for simplicity that all the users within the same cell transmit images of the same resolution and perform the video analytics utilizing the same neural network. Hence, it yields that the queuing system at the server can be modelled as an M/D/1 queue with load $\rho = \lambda T_{s}$.

\subsection{Video analytics}
All the servers in the network incorporate an object detection algorithm consisting of a pre-trained YOLOv5 neural network. We specifically consider YOLOv5 \cite{glenn2021yolov5} for the video analytics because it is capable of handling images of different resolutions without changing its associated learnable parameters, thus making it possible to parametrize the inference time and the accuracy of the detection algorithm.

Following the results in \cite{liu2018edge}, the number of instructions that are required to process an image with a resolution of $s_k^2$ pixels can be expressed as a convex function of $s_k$,
\begin{equation}
    f(s_k) =  c_1s_k^3 + c_2,
    \label{eq:nrof_instructions_YOLO}
\end{equation}
for some positive constants $c_1$ and $c_2$, where the units of $f(s_k)$ are number of floating-point operations (in trillions). Similarly, the accuracy of the detection algorithm can be expressed as a concave function of $s_k$,
\begin{equation}
    A(s_k) = c_3 - c_4e^{-c_5 s_k},
    \label{eq:accuracy_YOLO}
\end{equation}
for some other positive constants $c_3$, $c_4$, and $c_5$, where the accuracy of the detection is measured as the mean average precision of the object detection algorithm for a predefined threshold of the intersection over union \cite{liu2018edge}. For more information about the effect of the learning and inference processes on the constants $c_1,\dots,c_5$, please refer to \cite{bochkovskiy2020yolov4}.

%\VF{but are considered as unique in this work.} \JP{(I think this second part should go in the numerical results.)}

Moreover, all the computing servers in the network are equipped with a CPU of $h$ TFLOPs. With that, it follows from \eqref{eq:nrof_instructions_YOLO} that the time to process an image of resolution $s_k^2$ is
\begin{equation}
    T_{s,k} = \frac{f(s_k)}{h} = \frac{c_1s_k^3 + c_2}{h}.
    \label{eq:serviceTime}
\end{equation}
Note that images of high resolution take longer to process but they result in higher detection accuracies, with a nonlinear trade-off between the processing time and the accuracy of the detection algorithm.  %However, reducing the resolution of the images is not desirable when a small decrease in processing time comes at the expense of a large reduction of the classification accuracy.

\subsection{Offloading process}
\label{sec:offloading_process}
Altogether, the offloading process from beginning to end can be described as follows. 
%Text moved to earlier part: Within the Voronoi cell of each base station, we assume there is a set of $K$ active users, denoted by $k$, for $k=1,\dots,K$, that are capable of continuously recording images and sending them over the wireless channel. 
All active video analytics users are assigned some frequency resources and are considered to transmit with high priority, that is, they always get access to the wireless channel as soon as they want to offload their images. The transmitted images are considered to be squared with a resolution of $s_k^2$ pixels, encoded at $\theta$ bits per pixel and compressed at rate $\xi$:1, so an image of $s_k^2$ pixels is represented by $\theta s_k^2/\xi$ bits. Once the images are completely received at the serving base station, the images are queued at the edge server and processed one by one by a convolutional neural network. For each user $k$, we denote the time to transmit an image as $T_{ul,k}$, the time spent in the queue as $T_{w,k}$, and the time to process an image as $T_{s,k}$. We refer to the sum of these three times as the delay of the entire offloading process, disregarding the little time of eventual feedback to the user, and let each user $k$ specify a maximum value $t_k$ for that delay. For simplicity, we assume that the maximum delay $t_k$ is the same for all the users within the same cell.

\section{System analysis}
\label{sec:system_analysis}

\subsection{Coverage probability}
%\VF{I would still skip this first equation. It is completely unnecessary. We never use this in this unconditioned form, and therefore it is misleading.}

The coverage probability for any user $k$ in the network is the probability that its uplink SINR$_k$ is greater than some threshold $\gamma_k$.
% \begin{align}
%     p_{cov}&(\gamma_k, \lambda_b, \lambda_u, \delta, P, \alpha, \epsilon, \Bar{P}) = \mathcal{P}(\text{SINR}_k\geq\gamma_k)\nonumber\\
%     & = \int_{0}^{\infty} \mathcal{P}(\text{SINR}_k\geq\gamma_k|r) f_R(r) dr.
%     \label{eq:minimal_def_pCov}
% \end{align}
Considering this, the coverage probability conditioned to user $k$ being at a fixed distance $r$ from its serving base station can be derived from \cite[Theorem 1]{novlan2013analytical}, for any reuse factor $\delta$ \cite[Theorem 4]{andrews2011tractable}, and power-limit constraint \eqref{eq:power_constrain} as
%\JP{Then, if we perform the analysis on a randomly chosen user $k$ located at the origin and in distance $r$ from its serving base station, the conditional probability of coverage follows from \cite[Theorem 1]{novlan2013analytical}, for any reuse factor $\delta$ \cite[Theorem 4]{andrews2011tractable}, and power-limit constrain \eqref{eq:power_constrain} as}
%\VF{For a randomly chosen user $k$ in distance $r$ from its serving base station, the conditional coverage probability can be expressed by following the reasoning in \cite[Theorem 1]{novlan2013analytical}\cite[Theorem 4]{andrews2011tractable}, and considering the power-limit constraint in \eqref{eq:power_constrain}} 
%\TOFIX{Why reference \cite{andrews2011tractable}? Maybe we should state what is taken from these references.}
\begin{align}
    & \mathcal{P}(\text{SINR}_k\geq\gamma_k|r) \nonumber\\
    %& \; = \mathcal{P}\left(\frac{|g_k|^2 \,\ell(r,\alpha,\epsilon)\,r^{-\alpha}}{\sigma^2_k + \sum_{z\in\mathcal{Z}} |g_z|^2\, \ell(r_z,\alpha,\epsilon)\, d_z^{-\alpha}} \geq \gamma_k \bigg| r\right), \nonumber \\
    & \; = \mathcal{P}\left( |g_k|^2 \geq \frac{\gamma_k\, (\sigma_k^2 + \sum_{z\in\mathcal{Z}} |g_z|^2\, \ell(r_z,\alpha,\epsilon)\, d_z^{-\alpha})}{\ell(r,\alpha,\epsilon)\,r^{-\alpha}} \bigg| r\right) \nonumber \\
    %& \; \overset{(a)}{=} e^{- \frac{\gamma_k\, r^{\alpha}}{\ell(r,\alpha,\epsilon)}\, \sigma^2_k} \; \mathbb{E}_{r_z, g_z, d_z}\left[ \frac{\gamma_k\, r^{\alpha}}{\ell(r,\alpha,\epsilon)} \sum_{z\in\mathcal{Z}} |g_z|^2\, \ell(r_z,\alpha,\epsilon)\, d_z^{-\alpha} \right] \nonumber \\
    & \; \overset{(a)}{=} e^{- \frac{\gamma_k\, r^{\alpha}}{\ell(r,\alpha,\epsilon)}\, \sigma^2_k} \; \mathbb{E}\left[ \exp\left(-\frac{\gamma_k\, \sum_{z\in\mathcal{Z}} |g_z|^2\, \ell(r_z,\alpha,\epsilon)\, d_z^{-\alpha}}{\ell(r,\alpha,\epsilon)\,r^{-\alpha}} \right) \right] \nonumber \\
    & \; \overset{(b)}{=} e^{- \frac{\gamma_k\, r^{\alpha}}{\ell(r,\alpha,\epsilon)}\, \sigma^2_k} \mathcal{L}_{I_z}\left(\frac{\gamma_k\, r^{\alpha}}{\ell(r,\alpha,\epsilon)}\right), \label{eq:SINRk_givenR}
\end{align}
where $(a)$ follows from the exponential distribution of $|g_k|^2$, and $(b)$ follows from the definition of the Laplace transform of the interference at the serving base station. This Laplace transform can be calculated from the i.i.d.\@ distributions of $r_z$ and $d_z$, and the Probability Generating Function \cite{chiu2013stochastic} of the PPP that models the location of the mobile users, such that
\begin{equation}
    \mathcal{L}_{I_z}(s) = \exp\left(-2\pi\lambda_u\int_{r}^{\infty} \beta(x, s)\, x\, dx  \right), 
    \label{eq:laplaceTransform_pCov}
\end{equation}
with 
\begin{equation*}
\beta(x, s) = 1-\int_{0}^{\infty}\frac{\pi\lambda_u\, e^{-\pi\lambda_u u}}{1+s \,\ell\left(u^{\frac{1}{2}},\alpha,\epsilon\right) x^{-\alpha} } \,du,
\end{equation*}
where we considered that the closest interferer to the serving base station of user $k$ is at least at distance $d_z\geq r$.

%For this latter part, we considered the net interference at the serving base station of user $k$ to be the sum of all the received powers from all the users located at a distance greater than $r$.} \TOFIX{We can discuss this part further in person to see why I would like to keep the last sentence. The general idea is that equation \eqref{eq:SINRk_givenR} is calculated by considering that user $k$ is located at the origin, and the Laplace transform is calculated by considering that the BS is located at the origin. This is the conventional way of finding $p_{cov}$ but it needs to be clear what are the things that are separated a distance $r$ in \eqref{eq:SINRk_givenR} and what are the things that are separated in distance $r$ for the Laplace transform.} % For this latter part, we considered the net interference at the serving base station of user $k$ to be the sum of all the received powers from all the users located at a distance greater than $r$.
%\VF{We also need to consider that all interfering user has to be in a distance larger than $r$.}

%Note that the density of users always satisfy $\lambda_u\leq \lambda_b/\delta$ for our system model because there cannot be more active users than there are available frequency resources at the base stations.
\newpage
Note that the density of users using the same frequency resources always satisfies $\lambda_u\leq \lambda_b/\delta$ because there cannot be more active users than available frequency resources at the base stations.

\subsection{Ergodic capacity}
From the estimates of the state of the channel, any active user $k$ can adjust its coding and modulation according to its instantaneous SINR$_k$ to achieve a transmission rate close to the Shannon capacity, $B_k\log_2(1+\text{SINR}_k)$. Considering this, and following the same steps as the ones considered to find \eqref{eq:SINRk_givenR}, we can calculate the ergodic capacity for any user $k$ at distance $r$ from its serving base station as \cite[Theorem 3]{andrews2011tractable}
%\VF{can be calculated as} as}\sout{we can find the exact mathematical expression for} the ergodic capacity for a random chosen user $k$ located at a distance $r$ from its serving base station \VF{can be calculated as} as \cite[Theorem 3]{andrews2011tractable}
\begin{align}
    \Bar{R}_k|r & = \mathbb{E}[B_k \log_2(1+\text{SINR}_k)|r] \nonumber \\
    %& = \frac{B_k}{\log(2)}\mathbb{E}[\log(1+\text{SINR}_k)|r] \nonumber \\
    %& = \frac{B_k}{\log(2)} \int_{0}^{\infty} \mathcal{P}(\log(1+\text{SINR}_k)\geq t|r)dt \nonumber \\
    & = \frac{B_k}{\log(2)} \int_{0}^{\infty} \mathcal{P}(\text{SINR}_k\geq e^{t}-1|r)dt \nonumber \\
    & = \frac{B_k}{\log(2)} \int_{0}^{\infty}  e^{- \frac{(e^t-1)\, r^{\alpha}}{\ell(r,\alpha,\epsilon)}\sigma^2_k} \mathcal{L}_{I_z}\left(\frac{(e^t-1)\, r^{\alpha}}{\ell(r,\alpha,\epsilon)}\right)dt, \label{eq:ergodicCapacity_conditioned}
\end{align}
where the second equality follows from the fact that the expectation of any positive random variable $X$ can be calculated as $\mathbb{E}[X]=\int_0^\infty \mathcal{P}\left(X\geq t\right)dt$, and the last equality follows from the result in \eqref{eq:SINRk_givenR} by substituting $\gamma_k$ for $e^t-1$.

\subsection{Probability of successful computation}
The probability of successful computation measures the probability that the entire offloading process is completed within the maximum delay requirement $t_k$. We consider that the offloading process starts when a user begins offloading an image and ends when the server finishes processing that image. Hence, for a user $k$ at distance $r$ from its serving base station and for a maximum delay requirement $t_k$, the probability of successful computation can be expressed as
\begin{align}
    p_{succ}&\left(t_k, s_k, \lambda, \lambda_b, \lambda_u, \delta, P, \alpha, \epsilon, \Bar{P}\,|\,r\right) = \nonumber \\
    & \mathcal{P}\left(T_{ul,k}+T_{w,k}+T_{s,k}\leq t_k\,|\,r\right). \label{eq:probSuccessfulComputation}
\end{align}
Note that $T_{ul,k}$ is a random variable that depends on the sources of randomness in SINR$_k$, $T_{w,k}$ is a random variable that depends on the number of images that are already waiting in queue, and $T_{s,k}$ is a deterministic variable that depends on the image resolution $s_k^2$ according to \eqref{eq:serviceTime}. 
%Considering this, we can calculate the probability of successful computation by applying the Bayes' rule of conditional probability and finding the PDF of $T_{ul,k}|r$ and $T_{w,k}$.
Considering this, we can calculate \eqref{eq:probSuccessfulComputation} by applying the Bayes' rule of conditional probability and finding the PDF of $T_{ul,k}|r$ and $T_{w,k}$.

%\sout{Considering this, and starting from the definition in \eqref{eq:probSuccessfulComputation}, we can find the exact mathematical expression of the probability of successful computation by applying the Bayes' rule of conditional probability and finding the probability density functions of $T_{ul,k}|r$ and $T_{w,k}$.} \JP{(I think this text is necessary because otherwise it is not clear why we are suddenly computing the conditional PDF instead of just the PDF, but it can be shortened a bit).}

The PDF of $T_{ul,k}|r$ can be derived by considering that an image is completely offloaded at the base station when all its bits arrive at the base station. That is, when the sum of all the bits transmitted over $\left\lceil N_{\textit{coh}} \right\rceil$ coherence intervals of length $\tau_c$ is equal to (or larger than) the size of the image\footnote{For simplicity, we neglect the pilot signals, headers, and cyclic prefixes.},
\begin{equation}
    \tau_c\sum_{n=1}^{\left\lceil N_{\textit{coh}} \right\rceil} \log_2(1+\text{SINR}_k[n])|r \geq \frac{\theta s^2_k}{\xi},
    \label{eq:sum_dataRates}
\end{equation}
where $\left\lceil \cdot \right\rceil$ is the least integer function. While $\tau_c$ depends on the propagation environment, sending images with resolution $s_k^2\geq 40$ kilopixels require sufficiently large values of $N_{\textit{coh}}$ for the channel to experience a high variety of fading realizations during the entire uplink transmission \cite{torres2021lower}. Considering this, the left hand side of \eqref{eq:sum_dataRates} can be approximated as 
\begin{gather*}
    \tau_c \sum_{n=1}^{\left\lceil N_{\textit{coh}} \right\rceil} \log_2(1+\text{SINR}_k[n])|r \xrightarrow{N_{\textit{coh}}\rightarrow\infty} \\
    \tau_c \left\lceil N_{\textit{coh}} \right\rceil\,\mathbb{E}[\log_2(1+\text{SINR}_k)|r] = \frac{\tau_c \left\lceil N_{\textit{coh}} \right\rceil }{B_k} \, \Bar{R}_k|r,
\end{gather*}
where $\Bar{R}_k|r$ is the conditional ergodic capacity given in \eqref{eq:ergodicCapacity_conditioned}. Since $N_\textit{coh}=T_{ul,k}B_k/\tau_c$, we get
%
%\JP{Considering this, we can approximate \eqref{eq:sum_dataRates} to} 
%\sout{From a mathematical point of view, it can be derived from \cite{torres2021lower} that sending images with resolution $s_k^2\geq 40$ kilopixels require sufficiently large values of $N_{\textit{coh}}$ for the channel to experience different fading realizations during the entire uplink transmission. In such case,}
%
%\begin{gather*}
%    \tau_c \sum_{n=1}^{N_{\textit{coh}}} \log_2(1+\text{SINR}_k[n])|r \xrightarrow{N_{\textit{coh}}\rightarrow\infty} \\
%    \tau_c \, N_{\textit{coh}}\,\mathbb{E}[\log_2(1+\text{SINR}_k)|r] \geq \frac{\theta s^2_k}{\xi}.
%\end{gather*}
%Then, the minimum uplink transmission time follows from the definition of $\Bar{R}_k|r$ in \eqref{eq:ergodicCapacity_conditioned} and the fact that $N_\textit{coh}$ is calculated over a time-frequency block that expands over the entire uplink transmission and the frequency resources assigned to each user $k$, i.e., $N_\textit{coh}=T_{ul,k}B_k/\tau_c$, and
%
\begin{equation*}
   %\tau_c\, N_{\textit{coh}}\; \mathbb{E}[\log_2(1+\text{SINR})|r] \geq \frac{\theta s^2_k}{\xi}\; \Leftrightarrow \; 
   T_{ul,k}|r \geq \frac{\theta s^2_k}{\xi} \frac{1}{\Bar{R}_k|r}.
\end{equation*}
Note that the minimum time to transmit an image of $\theta s_k^2/\xi$ bits becomes a deterministic variable as $N_{\textit{coh}}\rightarrow\infty$. Thus, in the limit, we can express the PDF of $T_{ul,k}|r$ for a user $k$ located at distance $r$ from it serving base station as
% \begin{equation*}
%     \mathcal{P}(T_{ul,k}\leq\omega|r)\xrightarrow{N_{\textit{coh}}\rightarrow\infty}
%     \begin{cases}
%     1, & \textit{if }\; \omega \geq \frac{\theta s^2_k}{\xi\,\Bar{R}_k|r},\\
%     0, & \textit{otherwise}.
%     \end{cases}
% \end{equation*}
% from which the probability density function is
\begin{equation*}
    f_{T_{ul,k}|r}(\omega) \xrightarrow{N_{\textit{coh}}\rightarrow\infty} \Delta\left(\omega-\frac{\theta s^2_k}{\xi\,\Bar{R}_k|r}\right),
\end{equation*}
where $\Delta(\cdot)$ represents the Dirac delta function.

The probability density function of $T_{w,k}$ can be derived from the state probabilities of an M/D/1 queuing system using the Erlang’s principle of statistical equilibrium \cite[Sections 10.4.2 and 10.4.4]{iversen2010teletraffic}. For a detailed explanation on how to derive the PDF of $T_{w,k}$, please refer to \cite{iversen2010teletraffic} and the equations therein.

Putting all together, we can express the conditional probability of successful computation as
\begin{align}
    & p_{succ} \left(t_k, s_k,\lambda, \lambda_b, \lambda_u, \delta, P, \alpha, \epsilon, \Bar{P} \;|\; r\right) \nonumber \\ 
    & \; = \int_{0}^{\infty} \mathcal{P}(T_{w,k}+T_{s,k}\leq t_k - \omega | T_{ul,k} = \omega, r) f_{ T_{ul,k}|r}(\omega) d\omega \nonumber \\
    %& \; = \mathcal{P}\left(T_{w,k}+T_{s,k}\leq t_k - \frac{\theta s^2_k}{\xi\,\Bar{R}_k|r}\right)\\
    & \; = \mathcal{P}\left(T_{w,k}\leq t_k - \frac{\theta s^2_k}{\xi\,\Bar{R}_k|r} - \frac{c_1s_k^3+c_2}{h}\right), \label{eq:probSuccessfulComputation_complete}
\end{align}
which can be efficiently and accurately calculated by using a recursive formula based on Fry's equations of state \cite{iversen1999waiting}.

\subsection{Effective arrival rate}
Depending on the system parameters, the number of images offloaded to the server do not always correspond to the number of images that are successfully transmitted or processed within the delay requirement $t_k$. Considering this, we calculate the effective arrival rate of all the users connected to the same server as
\begin{equation}
    \lambda_{\textit{eff}} = \lambda \int_{0}^\infty p_{succ}\left(t_k, s_k, \lambda, \lambda_b, \lambda_u, \delta, P, \alpha, \epsilon, \Bar{P}\,|\,r\right) f_R(r) dr,
    \label{eq:effective_arrivalRate}
\end{equation}
and the maximum effective arrival rate with respect to $\lambda$ as
\begin{equation}
    \lambda^*_{\textit{eff}} \left(t_k, s_k, \lambda_b, \lambda_u, \delta, P, \alpha, \epsilon, \Bar{P}\right) = \max_{\lambda} \lambda_{\textit{eff}}.
    \label{eq:maxEffective_arrivalRate}
\end{equation}

% \subsection{Effective arrival rate \JP{TODO}}
% The results from Figs.\@ \ref{fig:pSuccRes_eps025_lambda025_refP10_final} and \ref{fig:pSuccRes_eps025_lambda025_refP10_final2} indicate that the images are not always successfully transmitted or processed within the delay requirement $t_k$. Considering this, we calculate the effective arrival rate of all the users connected to the same server as
% \begin{equation*}
%     \lambda_{\textit{eff}} = \lambda \int_{0}^\infty p_{succ}\left(t_k, s_k, \lambda, \lambda_b, \lambda_u, \delta, P, \alpha, \epsilon, \Bar{P}\,|\,r\right) f_R(r) dr,
% \end{equation*}
% and the maximum effective arrival rate with respect to $\lambda$ as
% \begin{equation*}
%     \lambda^*_{\textit{eff}} \left(t_k, s_k, \lambda_b, \lambda_u, \delta, P, \alpha, \epsilon, \Bar{P}\right) = \max_{\lambda} \lambda_{\textit{eff}}.
% \end{equation*}

\section{Numerical results}
\label{sec:numerical_results}
\begin{table}[t]
    \centering
    \caption{System parameters}
    \begin{tabular}{|l|c|}
        \hline
        Path-loss exponent $\alpha$ & 3.7 \\ \hline
        Peak transmission power $\Bar{P}$ & 23 dBm (200 mW) \\ \hline
        Noise power density $\sigma^2$ & -174 dBm/Hz \\ \hline
        Bandwidth assigned to each user $B_k$ & 2.1 MHz \\ \hline
        Image height/width $s_k$ & $[200,600]$ pixels \\ \hline
        Image encoding rate $\theta$ & 24 bits/pixel \\ \hline
        Image compression rate $\xi$ & 2 \\ \hline
        Maximum load at the server $\rho_{\max}$ & $0.99$ Erlangs\\ \hline
        Processing capabilities of the server $h$ & 10 TFLOPs \\ \hline
    \end{tabular}
    \label{tab:system_parameters_simulation}
\end{table}
%This section evaluates our system model in the context of realistic parameters for multi-cell, edge-intelligent systems supporting video analytics. 
% , 
This section evaluates all the performance metrics presented in Section \ref{sec:system_analysis} in the context of realistic parameters for multi-cell, edge-intelligent systems supporting video analytics. We also analyze and discuss the effect of the system parameters on the accuracy, effective frame rate, and system fairness.

In all our numerical evaluations, we consider the system parameters from Table \ref{tab:system_parameters_simulation}, corresponding to a typical cellular network \cite{ghosh2010fundamentals}.
%\TOFIX{I do not think it is transmission rate....Whose rate? It has to be something network wide.... Spectrum utilization? Network throughput?}
Following the same reasoning as \cite{andrews2011tractable}, we set $\delta=1$ to maximize the average transmission rate in the network. Moreover, we set the density of base stations to $\lambda_b=0.25$ BS/km$^2$, and consider the density of users using the same frequency resources to be $\lambda_u = \lambda_b$, corresponding to a highly loaded network. 
%\sout{such that the inter-cell interference is maximum. Moreover, following the same reasoning as in \cite{andrews2011tractable}, we consider that any increase in coverage probability from using a frequency reuse factor $\delta>1$ comes at the expense of reducing the available frequency resources in each cell by a factor $1/\delta$, so we set $\delta=1$ for maximizing the transmission rate.}
For the detection algorithm, we set the parameters that define \eqref{eq:nrof_instructions_YOLO} and \eqref{eq:accuracy_YOLO} to $c_1 = 7\cdot 10^{-10}$, $c_2=0.083$, $c_3=1$, $c_4=1.578$, and $c_5=6.5\cdot 10^{-3}$, which have been shown to fit the ground truth with a root mean square error less than $0.03$ \cite{liu2018edge}. Recall that we assumed for simplicity that the image resolutions $s_k^2$ and the maximum delay requirements $t_k$ are the same for all the users located within the same cell.

\begin{figure*}[!t]
    \centering
    \subfigure[]{\includegraphics[width=0.44\textwidth]{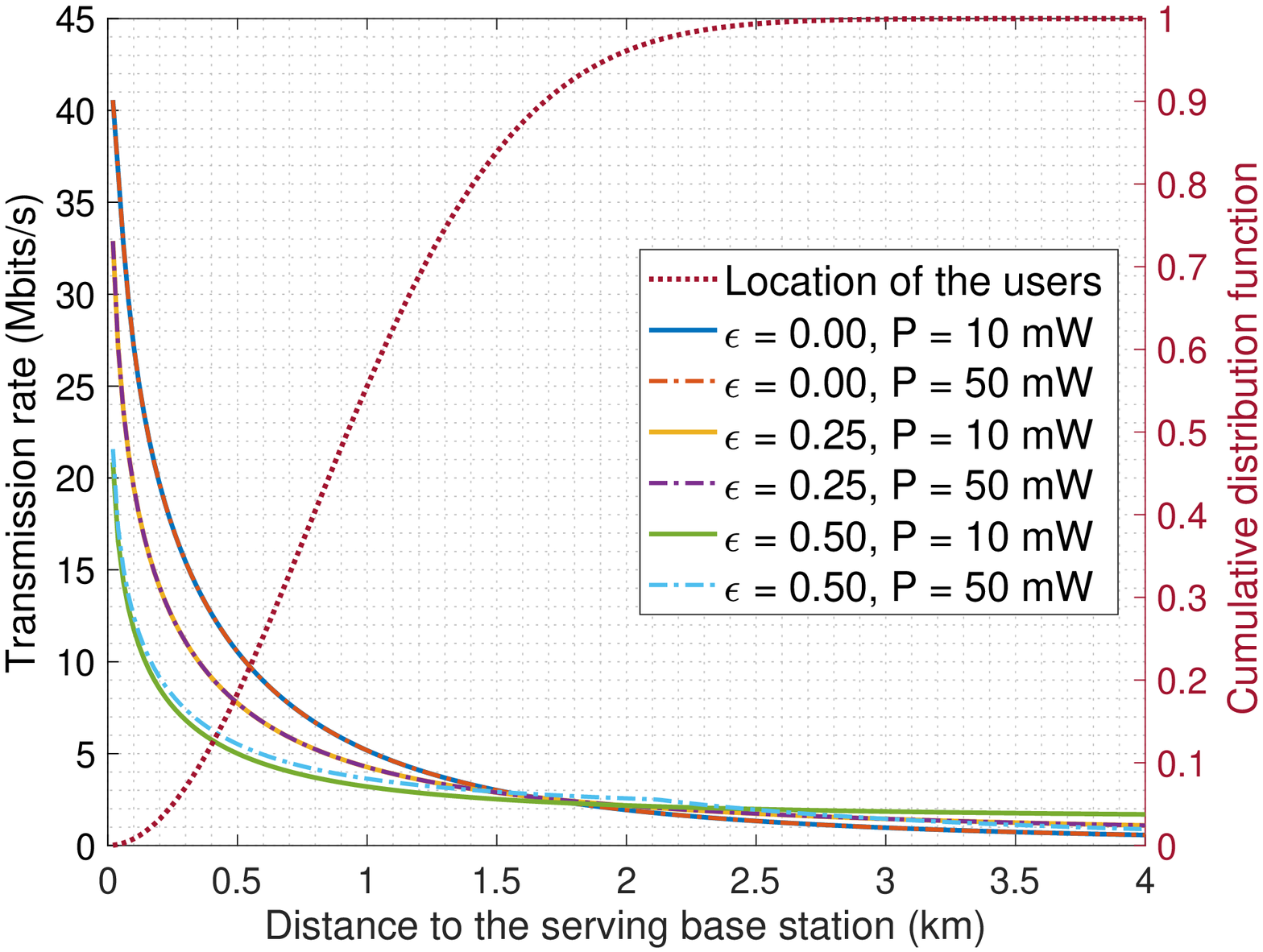} \label{fig:plotRate_refPower_B21_CDFRayleigh_final}}
    \subfigure[]{\includegraphics[width=0.44\textwidth]{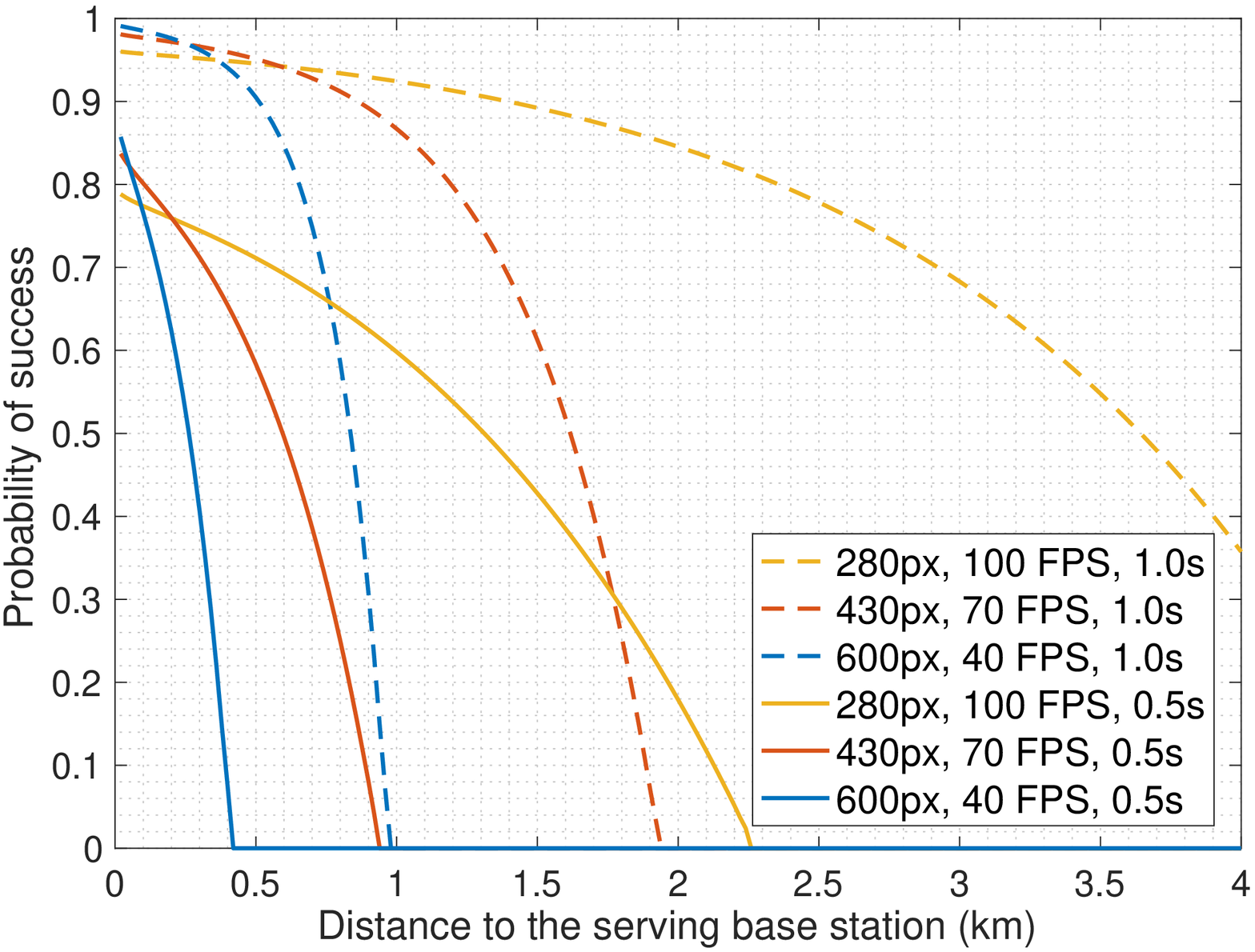} \label{fig:pSuccRes_eps025_lambda025_refP10_final}}
    \subfigure[]{\includegraphics[width=0.44\textwidth]{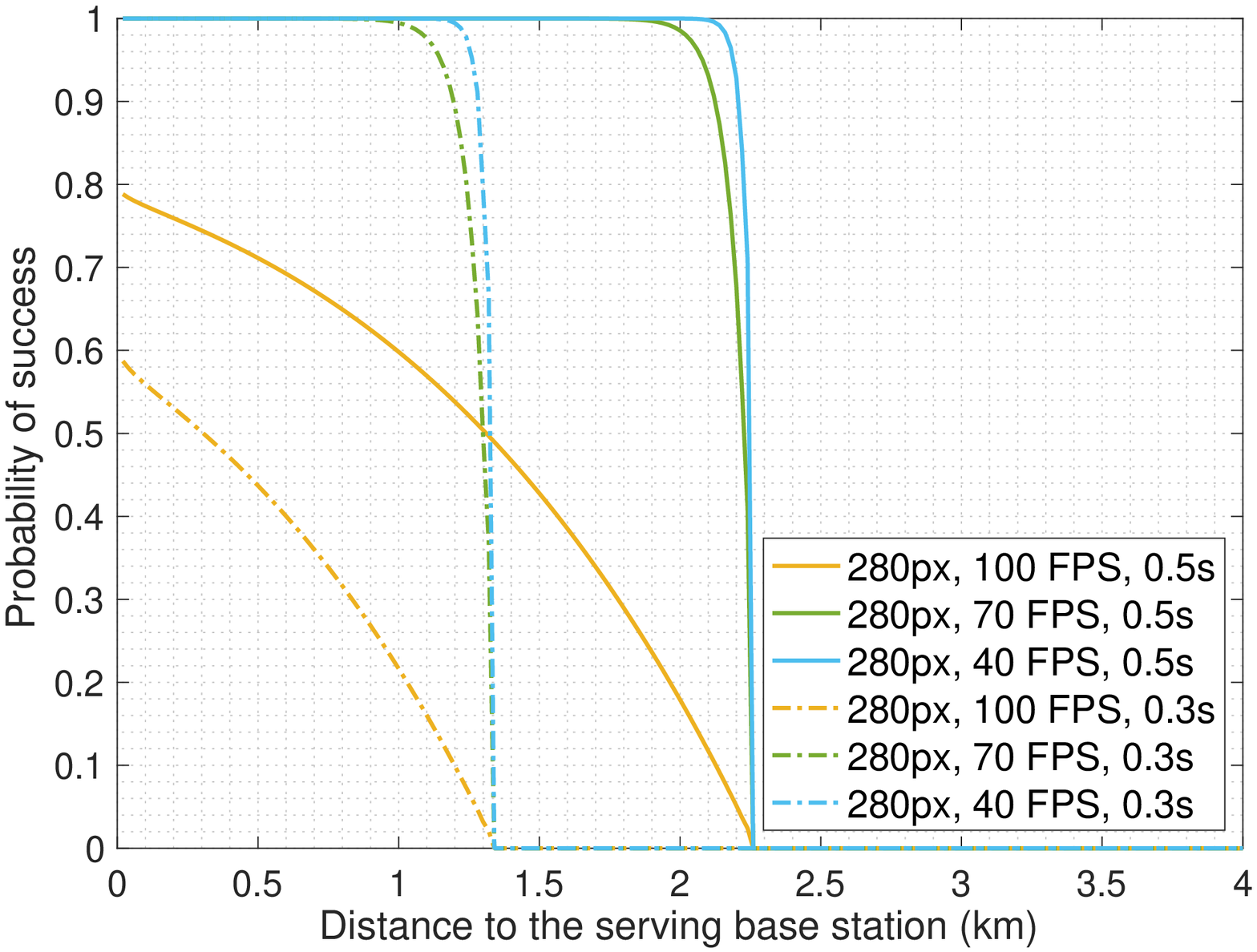} \label{fig:pSuccRes_eps025_lambda025_refP10_final2}}
    \subfigure[]{\includegraphics[width=0.44\textwidth]{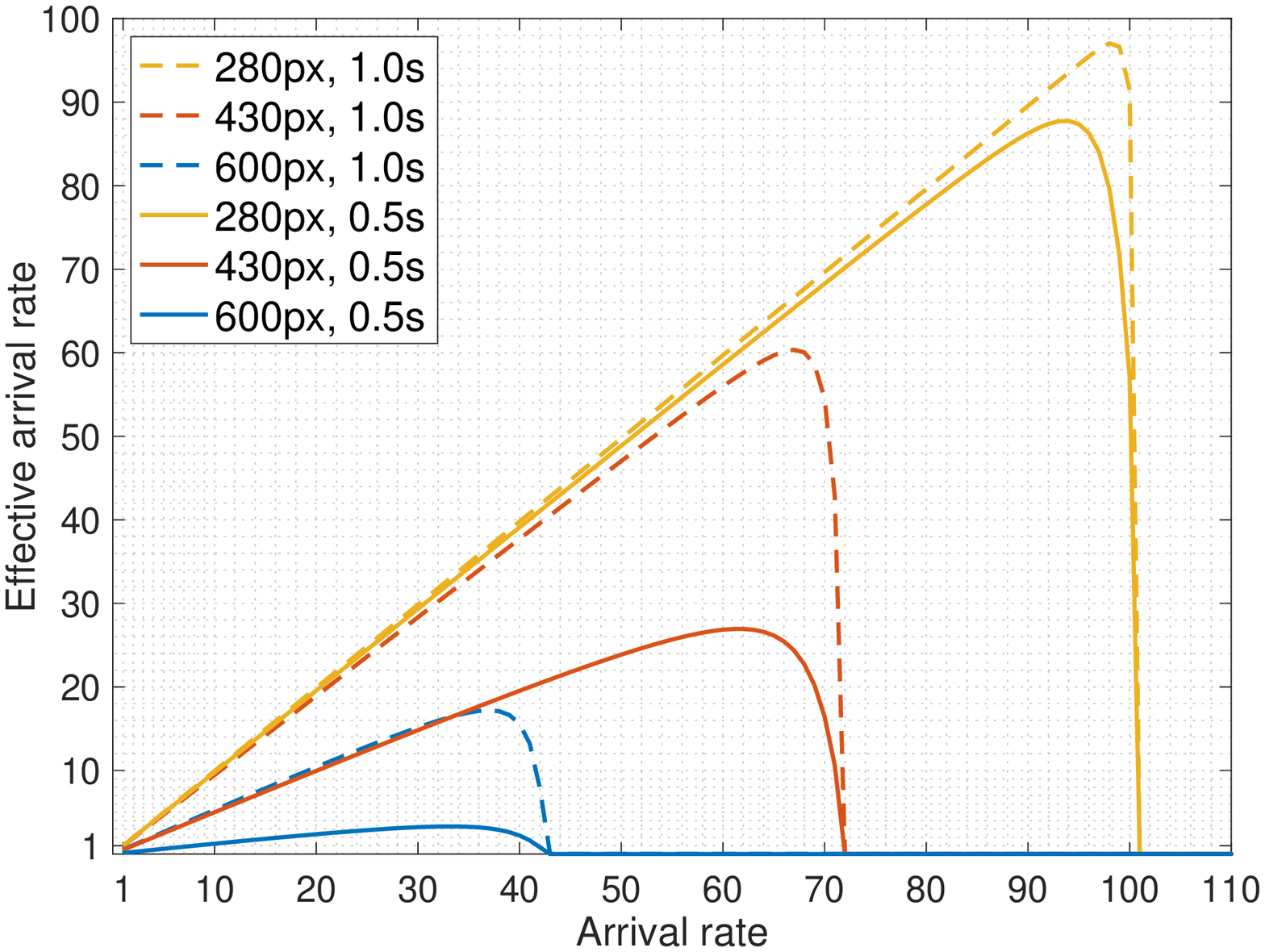} \label{fig:avgArrivalRate_eps025_lambda025_refP10_final}}
    \subfigure[]{\includegraphics[width=0.44\textwidth]{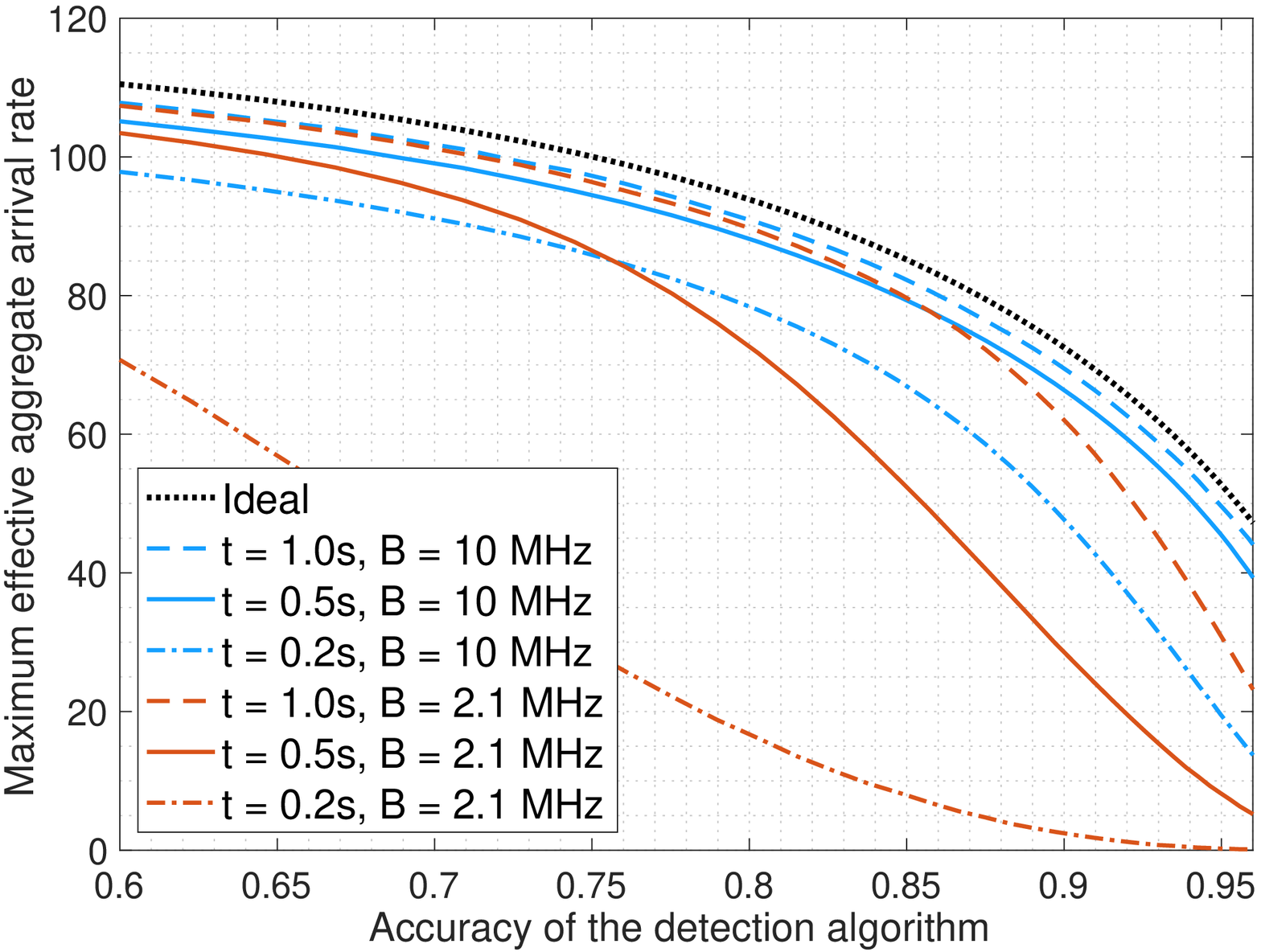} \label{fig:maxAvgEffectiveArrivalRate_accuracy_eps025_lambda025_refP10_final}}
    \subfigure[]{\includegraphics[width=0.44\textwidth]{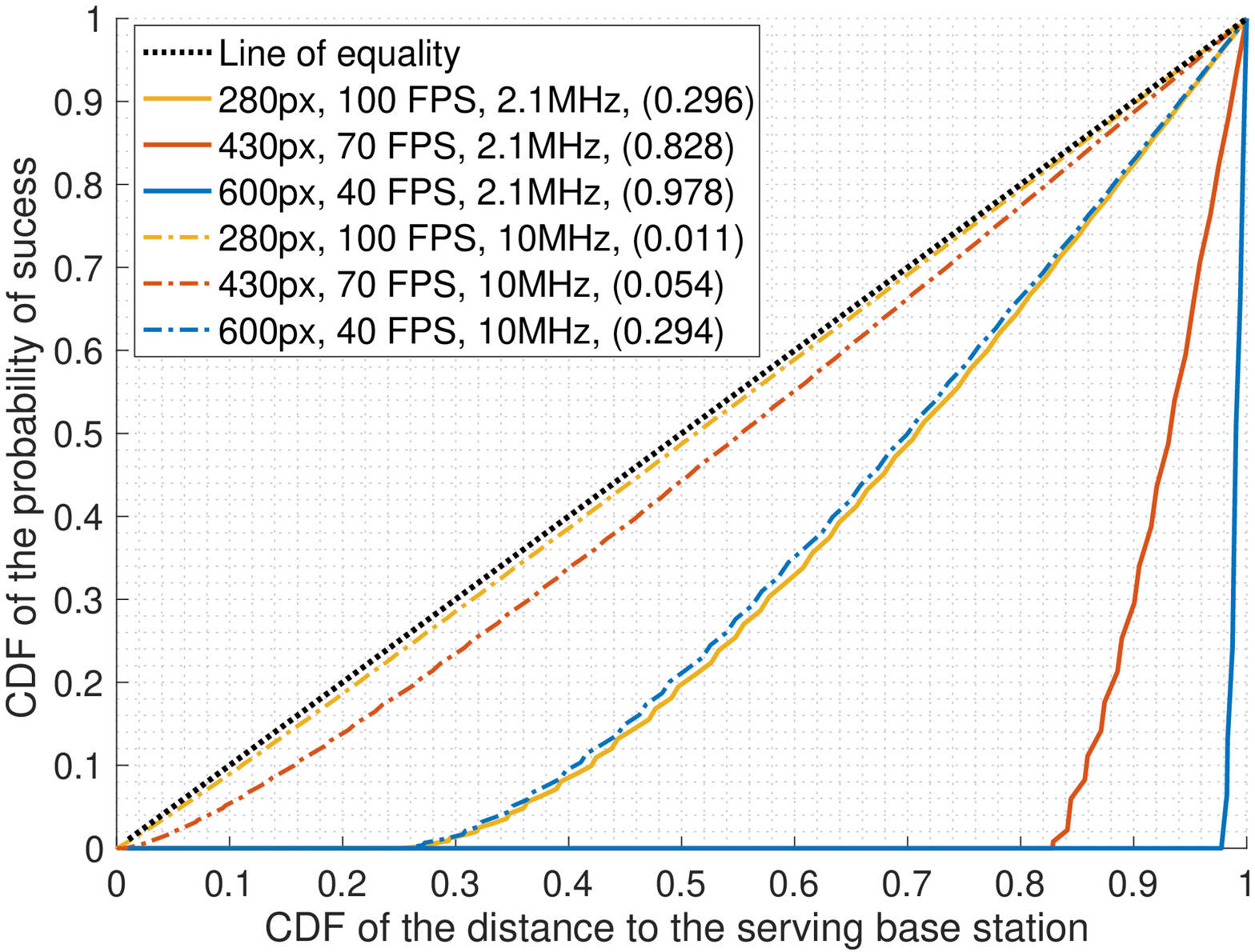} \label{fig:LorenzCurve_final_both}}
    \caption{(a)-(c) Conditional transmission rate and probability of successful computation as a function of the distance between any randomly selected user and its serving base station. (d) Effective aggregate arrival rate as a function of the arrival rate. (e) Maximum effective aggregate arrival rate as a function of the accuracy of the detection algorithm. (f) Lorenz curve with Lorenz curve gap given in the legend. Figures (a)-(c) evaluate the relationship between the different parameters of the video analytics, of the uplink transmission, and of the edge server. Figures (d) and (e) analyze the effect of the system parameters on the effective arrival rate and on the accuracy of the detection algorithm. Finally, Figure (f) measures the disparity of the probability that an image is successfully processed,
    depending on the location of the users within the coverage area of their serving base station.}
    \label{fig:simulation_results}
\end{figure*}

Figure \ref{fig:plotRate_refPower_B21_CDFRayleigh_final} shows the conditional transmission rate \eqref{eq:ergodicCapacity_conditioned} as a function of the distance $r$ for different fractional power control parameters $\epsilon$, and reference powers $P$. For all cases, the transmission rate decays rapidly with increasing $r$, and increasing $\epsilon$ leads to lower transmission rates for the users close to the base station but to slightly higher transmission rates for the users at the cell edge. Moreover, since we have an interference-limited system, $P$ makes little difference. %\TOFIX{I have not seen any paper that talks about conditional transmission rates, all the numerical results from all Andrews' papers focus on the coverage probability or the average transmission rate. Also, the $\epsilon$ that leads to the fastest decaying rate is $\epsilon=0$, and the only advantage of increasing $\epsilon$ is to provide better rates to the users at the cell-edge.}
The figure also shows the cumulative distribution function of the locations of the users within the coverage area of a serving base station, demonstrating that most of the users are located within the region from $0.5$ to $1.5$ km. To balance between supporting the users at the cell edge and providing a high transmission rate for the typical users, the rest of the numerical results are calculated with $\epsilon=0.25$ and $P=10$ mW.

%The left y-axis of figure \ref{fig:plotRate_refPower_B21_CDFRayleigh_final} shows the conditional transmission rate $\Bar{R}_k|r$ for different distances $r$, different fractional power controls $\epsilon$, and different reference powers $P$. The right y-axis of figure \ref{fig:plotRate_refPower_B21_CDFRayleigh_final} shows the cumulative distribution function of the locations of the users within the coverage area of a serving base station. In all cases, the users closer to the base station transmit at higher rates than the users further from the base station, but the transmission rate decays much faster as a function of the distance when all users transmit at the same power $(\epsilon=0)$ than when there is power control $(\epsilon>0)$. Moreover, for $\epsilon>0$, $55\%$ of the users (those located within a radius of $1$ kilometer) transmit at a power lower than $P$, but $\epsilon = 0.25$ is preferred over $\epsilon=0.5$ because it results in higher transmission rates for $95\%$ of the users (those located within a radius of $1.96$ kilometers). Considering this, the rest of the numerical results are calculated with $\epsilon=0.25$.

Figures \ref{fig:pSuccRes_eps025_lambda025_refP10_final} and \ref{fig:pSuccRes_eps025_lambda025_refP10_final2} show the conditional probability of successful computation \eqref{eq:probSuccessfulComputation_complete} as a function of the distance $r$ for different image heights/widths $s_k$, aggregate arrival rates $\lambda$, and delay requirements $t_k$. Specifically, the image sizes $s_k=\{280,430,600\}$ correspond to the maximum accuracy $A(s_k) = \{0.74, 0.90, 0.97\}$ that can be supported for the arrival rates $\lambda=\{100, 70, 40\}$, respectively, without overloading the servers. 
%\VF{Fig.\@ \ref{fig:pSuccRes_eps025_lambda025_refP10_final} shows that for users close to the base station the server load and the delay limit $t_k$ determines the success probability. Users further away for the base station are instead limited by the time it takes to transmit the image, and the success probability is strongly affected by the image size. Fig. \ref{fig:pSuccRes_eps025_lambda025_refP10_final2} shows that the image size and the delay limit determines the maximum distance of possibly successful image analytics. Within this distance $p_{succ}|r$ can be increased by decreasing the arrival rate, but it does not help the users far away from the base station.}
Figure \ref{fig:pSuccRes_eps025_lambda025_refP10_final} shows that the server load and the delay limit $t_k$ determines the success probability for the users close to the base station. Users farther away from the base station are instead limited by the uplink transmission time, and therefore the success probability is strongly affected by the image size. Figure \ref{fig:pSuccRes_eps025_lambda025_refP10_final2} shows that the image size and the delay limit determines the maximum distance of possibly successful image analytics. Within this distance, the probability of success can be increased by decreasing the aggregate arrival rate, but it does not help the users farther away from the base station.
%
% Note that the users are affected in different ways depending on the selected set of the parameters. When $s_k$ is large, only those users that are close to the base station can transmit their images within $t_k$, and the waiting time is dictated by the time to process the images. When $s_k$ is small, more users can transmit their images within $t_k$, and the waiting time is dictated by the amount of images received at the server. Overall, $s_k$ determines the maximum distance $r$ for which $p_{succ}|r>0$, $\lambda$ determines the decaying shape of $p_{succ}|r$, and $t_k$ shifts the graph along the x-axis.
%

Figure \ref{fig:avgArrivalRate_eps025_lambda025_refP10_final} shows the effective aggregate arrival rate \eqref{eq:effective_arrivalRate} as a function of the aggregate arrival rate for different $s_k$ and $t_k$. In all cases, we observe two distinct regions. In the first region, $\lambda_{\textit{eff}}$ grows linearly with $\lambda$, with shallower slopes for larger images sizes and more stringent delay requirements.
%\VF{In this region, the loss is due to the long transmission times of users far from the base station.} In the second region, $\lambda_{\textit{eff}}$ reaches its peak and then falls rapidly, \VF{as the server becomes overloaded}.
In this case, the loss is due to the long transmission times of the users farther from the base station. In the second region, $\lambda_{\textit{eff}}$ reaches its peak and then falls rapidly, as the server becomes overloaded.

% These figures
Figures \ref{fig:pSuccRes_eps025_lambda025_refP10_final}--\ref{fig:avgArrivalRate_eps025_lambda025_refP10_final} reflect that the system can be bandwidth-limited or computationally-limited, and it is important to know in which region the system operates to effectively decrease the losses.
%These results give insight on the selection of $\lambda$, as it is preferable to operate at an arrival rate slightly lower than the maximum achievable $\lambda$ for maximizing the number of images that are successfully transmitted and processed within the delay requirement $t_k$.
%Figures \ref{fig:maxAvgEffectiveArrivalRate_accuracy_eps025_lambda025_refP10_final} and \ref{fig:LorenzCurve_final_both} summarize
Now we continue with the analysis of the performance and the fairness of multi-cell edge video analytics.

Figure \ref{fig:maxAvgEffectiveArrivalRate_accuracy_eps025_lambda025_refP10_final} shows the maximum effective arrival rate \eqref{eq:maxEffective_arrivalRate} -- detection accuracy \eqref{eq:accuracy_YOLO} region of the system. We consider $B_k = 2.1$ MHz and $B_k=10$ MHz to distinguish between the bandwidth-limited and the computationally-limited cases. The results are compared to the ideal case with maximum server load and no losses. We see that the rate-accuracy trade-off of the computationally-limited systems follows the shape of the ideal case, with decreasing values under stricter deadlines $t_k$. However, the bandwidth-limited systems behave differently, with an early decrease of the achievable rate as the accuracy is increased. Thus, the bandwidth-limited systems have little possibility to trade frame rate for increased accuracy.

Finally, to evaluate the fairness of the system, Figure \ref{fig:LorenzCurve_final_both} shows the Lorenz curve of the probability of successful computation  \eqref{eq:probSuccessfulComputation_complete} with respect to the location of the users, 
for different $s_k$, $\lambda$, $B_k$, and $t_k=0.3$ seconds. The Lorenz curve gap is given in the legend.
%for each configuration.
For all the configurations,
%the load is close to the maximum possible for stable systems.
we select the values of $s_k$ and $\lambda$ such that the load at server is close to $\rho_{\max}$.
We see that the computationally-limited systems (large bandwidth, smaller image size, and high arrival rate) remain fair, even though the probability of success is not necessarily high. On the other hand, there are bandwidth-limited systems (small bandwidth, larger image size, and low arrival rate) that are highly unfair, with a Lorenz curve gap close to one. In the middle we find configurations that are equally limited by the transmission and the computation capacities, where the success probability gradually decreases
%as the users distance to its base station is increased.
for increasing distances to the serving base station.
Overall, the system fairness can be improved by decreasing the frame size.%, which in turn allows for higher arrival intensities if one aims to maintain high server utilization.

\section{Discussion}
\label{sec:discussion}
Motivated by the recent advances in edge computing and artificial intelligence, we proposed a mathematical framework for modelling multi-cell, edge-intelligent systems supporting video analytics. We expressed the probability of successful computation and the effective arrival rate as a function of the accuracy of the video analytics, as well as the network and computing parameters. Our work provides the first steps towards designing efficient resource-allocation strategies and traffic-control protocols.
%It shows
We showed, for example, that different actions are necessary to improve the effective arrival rate, depending on whether the system is operating in bandwidth or computationally limited regimes.
%It also shows
We also showed that the system can be highly unfair, and thus centralized solutions may be necessary to protect the weaker users. Further work should consider the modelling of more flexible bandwidth allocation and scheduling solutions, as well as protocol design.

% By characterizing the system in the bandwidth and computationally limited regimes, 

% .. lets see how much space we have.}

% We derived expressions for the coverage probability, the transmission rate, the probability of successful computation, and the effective arrival rate. Our numerical results indicate that small-bandwidth systems are limited by the network resources, specially when users transmit large-size images, but the system can still remain fair if the users offload their images less frequently. Contrarily, high-bandwidth systems are limited by the computational resources at the server, but the users can select from a larger variety of configurations that lead to higher detection accuracies, higher arrival rates, and more fairness among other users within the same cell.

% This work is arguably the first to propose a mathematical model for edge intelligence in large-scale wireless systems. We will extend it in the future by considering more advanced queuing models that allow for the analysis of multiple users with different accuracy-latency requirements.

\balance
\bibliography{references}
\bibliographystyle{IEEEtran}

\end{document}